\documentclass[a4paper,fleqn,usenatbib]{mnras}

\usepackage{graphicx}	
\usepackage{amsmath,amsfonts,amssymb}
\usepackage{makeidx}
\usepackage{epstopdf} 
\usepackage{booktabs,caption,fixltx2e}
\usepackage[flushleft]{threeparttable}
\usepackage{hyperref}
\usepackage{multirow}
\usepackage{rotating}
\usepackage{color}
\usepackage[T1]{fontenc}
\usepackage{ae,aecompl}
\usepackage{soul}

\def \BE{\begin{equation}}
\def \EE{\end{equation}}	
\def \BC{\begin{center}}
\def \EC{\end{center}}
\def \BEA{\begin{eqnarray}}
\def \EEA{\end{eqnarray}}
\newcommand\planck{\textit{Planck}}
\newcommand\sptsz{SPT-SZ}
\newcommand\sptpol{SPTpol}
\newcommand\mnpsq{\ensuremath{\left< p^2\right>}}
\newcommand\psq{\ensuremath{ p^2}}
\newcommand\mnp{\ensuremath{\left< p\right>}}
\newcommand\sqdeg{\ensuremath{{\rm deg}^2}}
\newcommand\uksq{\ensuremath{\mu{\rm K}^2}}
\newcommand\Q{\ensuremath{Q}}
\newcommand\U{\ensuremath{U}}
\newcommand\I{\ensuremath{I}}
\newcommand{\comment}[1]{}

\newcommand{\ukarcmin}{\ensuremath{\mu{\rm K-arcmin}}}
\defcitealias{gupta17}{G17}
\defcitealias{everett19}{E19}

\newcommand{\Melbourne}{$^{1}$}
\newcommand{\Cardiff}{$^{2}$}
\newcommand{\FNAL}{$^{3}$}
\newcommand{\KICPChicago}{$^{4}$}
\newcommand{\illast}{$^{5}$}
\newcommand{\NIST}{$^{6}$}
\newcommand{\Berkeley}{$^{7}$}
\newcommand{\ArgonneHEP}{$^{8}$}
\newcommand{\AAUChicago}{$^{9}$}
\newcommand{\PhysicsUChicago}{$^{10}$}
\newcommand{\EFIChicago}{$^{11}$}
\newcommand{\UKZN}{$^{12}$}
\newcommand{\UChicago}{$^{13}$}
\newcommand{\TAPIRCaltech}{$^{14}$}
\newcommand{\Caltech}{$^{15}$}
\newcommand{\LBNL}{$^{16}$}
\newcommand{\McGill}{$^{17}$}
\newcommand{\CIFAR}{$^{18}$}
\newcommand{\ColoradoAPS}{$^{19}$}
\newcommand{\illphy}{$^{20}$}
\newcommand{\HarveyMudd}{$^{21}$}
\newcommand{\esogarching}{$^{22}$}
\newcommand{\ColoradoPhys}{$^{23}$}
\newcommand{\SLAC}{$^{24}$}
\newcommand{\Stanford}{$^{25}$}
\newcommand{\Davis}{$^{26}$}
\newcommand{\Arizona}{$^{27}$}
\newcommand{\Michigan}{$^{28}$}
\newcommand{\Munich}{$^{29}$}
\newcommand{\ExcellenceCluster}{$^{30}$}
\newcommand{\MPE}{$^{31}$}
\newcommand{\Dunlap}{$^{32}$}
\newcommand{\ArgonneMSD}{$^{33}$}
\newcommand{\Minnesota}{$^{34}$}
\newcommand{\CaseWestern}{$^{35}$}
\newcommand{\ArtInstChicago}{$^{36}$}
\newcommand{\ThreeSpeedLogic}{$^{37}$}
\newcommand{\JPL}{$^{38}$}
\newcommand{\CfA}{$^{39}$}
\newcommand{\KIPAC}{$^{40}$}
\newcommand{\GSFC}{$^{41}$}
\newcommand{\UToronto}{$^{42}$}
\newcommand{\Maryland}{$^{43}$}
\newcommand{\UCLA}{$^{44}$}

\author[N.~Gupta, C.~L.~Reichardt, et al.]{\parbox{\textwidth}{\Large 
  N.~Gupta\Melbourne\thanks{nikhel.gupta@unimelb.edu.au},
  C.~L.~Reichardt\Melbourne,
  P.~A.~R.~Ade\Cardiff,   
  A.~J.~Anderson\FNAL$^,$\KICPChicago,
  M.~Archipley\illast,
  J.~E.~Austermann\NIST,  
  J.~S.~Avva\Berkeley,  
  J.~A.~Beall\NIST,   
  A.~N.~Bender\ArgonneHEP$^,$\KICPChicago,
  B.~A.~Benson\KICPChicago$^,$\AAUChicago$^,$\FNAL,  
  F.~Bianchini\Melbourne,  
  L.~E.~Bleem\KICPChicago$^,$\ArgonneHEP,  
  J.~E.~Carlstrom\KICPChicago$^,$\AAUChicago$^,$\ArgonneHEP$^,$\PhysicsUChicago$^,$\EFIChicago,  
  C.~L.~Chang\KICPChicago$^,$\AAUChicago$^,$\ArgonneHEP,  
  H.~C.~Chiang\UKZN,  
  R.~Citron\UChicago,   
  C.~Corbett~Moran\TAPIRCaltech, 
  T.~M.~Crawford\KICPChicago$^,$\AAUChicago,  
  A.~T.~Crites\KICPChicago$^,$\AAUChicago$^,$\Caltech,  
  T.~de~Haan\Berkeley$^,$\LBNL, 
  M.~A.~Dobbs\McGill$^,$\CIFAR,  
  W.~Everett\ColoradoAPS,   
  C.~Feng\illast$^,$\illphy,
  J.~Gallicchio\KICPChicago$^,$\HarveyMudd,  
  E.~M.~George\Berkeley$^,$\esogarching,  
  A.~Gilbert\McGill,  
  N.~W.~Halverson\ColoradoAPS$^,$\ColoradoPhys,  
  N.~Harrington\Berkeley,  
  J.~W.~Henning\KICPChicago$^,$\ArgonneHEP,  
  G.~C.~Hilton\NIST,  
  G.~P.~Holder\CIFAR$^,$\illast$^,$\illphy,  
  W.~L.~Holzapfel\Berkeley,  
  Z.~Hou\KICPChicago,  
  J.~D.~Hrubes\UChicago,   
  N.~Huang\Berkeley,  
  J.~Hubmayr\NIST,  
  K.~D.~Irwin\SLAC$^,$\Stanford,  
  L.~Knox\Davis,  
  A.~T.~Lee\Berkeley$^,$\LBNL, 
  D.~Li\NIST$^,$\SLAC,  
  A.~Lowitz\UChicago, 
  D.~Luong-Van\UChicago,
  D.~P.~Marrone\Arizona,
  J.~J.~McMahon\Michigan,  
  S.~S.~Meyer\KICPChicago$^,$\AAUChicago$^,$\PhysicsUChicago$^,$\EFIChicago,   
  L.~M.~Mocanu\KICPChicago$^,$\AAUChicago,  
  J.~J.~Mohr\Munich$^,$\ExcellenceCluster$^,$\MPE,
  J.~Montgomery\McGill,
  A.~Nadolski\illast$^,$\illphy,  
  T.~Natoli\PhysicsUChicago$^,$\KICPChicago$^,$\Dunlap,  
  J.~P.~Nibarger\NIST,  
  G.~I.~Noble\McGill,  
  V.~Novosad\ArgonneMSD,  
  S.~Padin\KICPChicago$^,$\AAUChicago$^,$\Caltech,  
  S.~Patil\Melbourne,  
  C.~Pryke\Minnesota, 
  J.~E.~Ruhl\CaseWestern,  
  B.~R.~Saliwanchik\UKZN,  
  J.T.~Sayre\ColoradoAPS$^,$\ColoradoPhys,  
  K.~K.~Schaffer\KICPChicago$^,$\EFIChicago$^,$\ArtInstChicago,   
  E.~Shirokoff\Berkeley$^,$\KICPChicago$^,$\AAUChicago, 
  C.~Sievers\UChicago, 
  G.~Smecher\McGill$^,$\ThreeSpeedLogic,  
  Z.~Staniszewski\CaseWestern$^,$\JPL,
  A.~A.~Stark\CfA,  
  K.~T.~Story\Stanford$^,$\KIPAC,  
  E.~R.~Switzer\KICPChicago$^,$\GSFC,
  C.~Tucker\Cardiff, 
  K.~Vanderlinde\Dunlap$^,$\UToronto,   
  T.~Veach\Maryland,  
  J.~D.~Vieira\illast$^,$\illphy, 
  G.~Wang\ArgonneHEP,  
  N.~Whitehorn\UCLA,  
  R.~Williamson\UChicago$^,$\JPL,
  W.~L.~K.~Wu\KICPChicago, 
  V.~Yefremenko\ArgonneHEP,
  and
  L.~Zhang\illast
}
\vspace{0.4cm}
\\
\parbox{\textwidth}
{
The authors' affiliations are shown in the end of manuscript.
}
}

\begin{document}
\title[Fractional Polazisation of Extragalactic Radio Sources]{Fractional Polarisation of Extragalactic Sources in the 500-square-degree SPTpol Survey}

\maketitle

\date{Accepted ???. Received ???; in original form ???} 

\begin{abstract}
We study the polarisation properties of extragalactic sources at 95 and 150\,GHz in the SPTpol 500\,deg$^2$ survey. 
We estimate the polarised power by stacking maps at known source positions, and correct for noise bias by subtracting the mean polarised power at random positions in the maps. 
We show that the method is unbiased using a set of simulated maps with similar noise properties to the real  SPTpol maps. 
We find a flux-weighted mean-squared polarisation fraction $\langle p^2 \rangle=  [8.9\pm1.1] \times 10^{-4}$ at 95\,GHz and $[6.9\pm1.1] \times 10^{-4}$ at 150~GHz for the full sample. 
This is consistent with the values obtained for a sub-sample of active galactic nuclei. 
For dusty sources, we find 95\,per cent upper limits of $\mnpsq_{\rm 95}<16.9 \times 10^{-3}$ and $\mnpsq_{\rm 150}<2.6 \times 10^{-3}$.
We find no evidence that the polarisation fraction depends on the source flux or observing frequency. 
The 1-$\sigma$ upper limit on measured mean squared polarisation fraction at 150\,GHz implies that extragalactic foregrounds will be subdominant to the CMB E and B mode polarisation power spectra out to at least $\ell\lesssim5700$ ($\ell\lesssim4700$) and $\ell\lesssim5300$ ($\ell\lesssim3600$), respectively at 95 (150)\,GHz. 
\end{abstract}


\section{Introduction} 
\label{sec:intro}
Extragalactic sources at millimeter wavelengths can be classified into two broad categories: active galactic nuclei (AGN) and dust-enshrouded star-forming galaxies (DSFGs). 
While individual sources may have emission from both non-thermal and thermal emission, for AGN the emission is dominated by synchrotron radiation from the relativistic jets coming off the central black hole \citep[e.g.][]{best06,coble07,best12}. 
The signal from DSFGs is dominated by thermal dust emission \citep[e.g.][]{vieira10, tucci11}. 
These sources are well studied in temperature but the polarisation properties at millimeter wavelengths are less known. 
For both AGN and DSFGs, we expect some polarisation from interactions with magnetic fields. 
Thus polarised studies can inform us about the magnetic field structure of these objects. 
The focus of this paper, however, is to study the impact of the polarised emission from AGN and DSFGs on measurements of cosmic microwave background (CMB) polarisation at small angular scales. 

Extending measurements of CMB polarisation to smaller angular scales adds cosmological information \citep[e.g.][]{scott16}. 
With the exquisite sensitivity of upcoming CMB experiments like the Simons Observatory \citep{SO19} and CMB-S4 \citep{abazajian16}, the limiting factor on the angular scales used for cosmological analyses may be these polarised extragalactic foregrounds instead of instrumental noise.
Therefore it is important to understand the polarisation properties of these extragalactic sources in the key frequency bands for CMB science ($\sim$\,90 - 150\,GHz), and the resulting polarised foreground power at small angular scales. 

The polarisation properties of AGN are well-studied at radio frequencies, with a number of works finding polarisation fractions of a few per cent. 
 For instance, \citet{condon98} studied the polarisation of $\sim$\,30,000 radio sources in NVSS at 1.4\,GHz, and found a mean fractional polarisation $\langle p \rangle$ of 2 to 2.7 per cent. 
 Several authors have looked at AGN polarisation using ATCA data at 20\,GHz, finding numbers in the range of 2.3 - 4.8 per cent \citep{ricci04, sadler06, murphy10}. 
Data from VLA has been used to extend these measurements up to 43\,GHz, with \citet{sajina11} finding the mean polarisation of sources selected from the Australia Telescope 20 GHz (AT20G) survey with flux density $S_{\rm 20\,GHz} > 40$\,mJy to be in the range  of 2.5 to 5 per cent, but with some sources being up to 20 per cent polarised. 
It is unclear if these results will extend to the small subset of radio galaxies that are bright at 150\,GHz. 
\citet{galluzzi18} examined the frequency scaling of polarised emission across nearly 3 decades from 72\,MHz to 38\,GHz, and found the polarised spectra required the emission model to include more components than the intensity data. 
Thus while these works paint a consistent picture of AGN polarisation in the GHz to 10s of GHz range, extrapolating the results to the key CMB frequencies around 150\,GHz introduces significant uncertainty. 

In recent years, we have seen the first measurements of the polarisation properties of AGN at CMB frequencies, although these measurements have been restricted to the brightest AGN. 
Using data from the  {\it Planck} satellite, \citet{bonavera17} found $\langle p \rangle = 2.9^{+0.3}_{-0.5}$ per cent and \citet{trombetti18} found  $\langle p \rangle = 3.06\pm0.28$ per cent at 143~GHz for sources above 1\,Jy and 525\,mJy, respectively. 
An analysis of Atacama Cosmology Telescope (ACTpol) data found a consistent $\langle p \rangle = 2.8\pm0.5$ per cent for sources brighter than 215\,mJy at 148\,GHz \citep{datta18}. 
The brightest AGN are masked in CMB power-spectrum and lensing analyses. The DSFGs and AGN that remain will have fluxes $\lesssim$\,10\,mJy, much fainter than the sources in existing studies. 
The central goal of this work is to extend these measurements towards these lower flux sources.  

In this work, we present the first measurement of the polarisation properties of faint extragalactic sources (down to 6\,mJy at 150\,GHz) at CMB frequencies. 
The list of sources is drawn from the source catalog of the 2500\,\sqdeg{} \sptsz{} survey \citep[hereafter E19]{everett19}. 
The polarisation properties of these sources are measured using data from the 500\,\sqdeg{} \sptpol{} survey. 
We look at the mean polarisation properties, as well as the properties as a function of flux or frequency. 
Finally, we consider the impact of AGN and DSFG polarisation on measurements of CMB polarisation. 

The paper is structured as follows.
 In Section~\ref{sec:observations}, we describe the SPT-SZ point source catalogue and SPTpol maps. 
In Section~\ref{sec:method}, we describe and test the estimator on simulations. 
We present the measured polarisation fraction  in Section~\ref{sec:results}, and the implications for CMB polarisation measurements in Section~\ref{sec:power_spectrum}. 
Finally in Section~\ref{sec:conclusions}, we summarize our findings.

\section{Observations and Data Reduction}
\label{sec:observations}
This work uses temperature and polarisation data from the \sptpol{} survey to measure the polarisation properties of AGN and DSFGs in the \sptsz{} source catalogue. 
We briefly review both surveys here.  

\subsection{The SPT-SZ source catalogue and selection criteria}
\label{sec:catalog}
\citetalias{everett19} presents a catalogue of compact sources found in three-frequency data from the \sptsz{} survey, a 2500\,\sqdeg{} survey conducted using the 10-m South Pole Telescope \citep{carlstrom11}. 
In this work we measure the polarisation properties of a sub-sample of these sources. 
Here we review the catalogue and selection criteria for this work. 

Briefly, the source catalogue in \citetalias{everett19} was generated by applying a matched filter  \citep{tegmark98} to the \sptsz{} maps at each frequency, in order to optimise the signal-to-noise of beam-sized objects. 
The CLEAN algorithm \citep{hogbom74} was used to identify sets of bright pixels in the filtered map as individual objects, and to calculate the flux of each object. 
Sources were classified by cross-matching against other catalogues and by measuring the spectral indices from 95 to 150\,GHz and 150 to 220\,GHz. 

This work applies three selection criteria to the \citetalias{everett19} catalogue. 
First, we require the 150\,GHz flux to be $S_{\rm 150\,GHz} > 6$\,mJy, which corresponds to approximately a 5-$\sigma$ detection threshold. 
The purity rate in the sample above this flux is very high i.e. 90-98 per cent depending upon the noise at the source location.
Second, we require the sources to be compact and to not have a stellar counterpart. 
Finally, we restrict the list to sources within a 470.8\,\sqdeg{} region of the \sptpol{} survey with uniform noise.
These criteria leave a sample of 686 galaxies, of which 92~per cent are AGN and the rest are DSFGs. 

\subsection{The 500\,\sqdeg{} SPTpol survey}
\label{sec:maps}
The polarisation-sensitive \sptpol{} receiver was installed on the South Pole Telescope in the austral summer of 2011-2012. 
The receiver has 180 and 588 polarisation-sensitive pixels at 95 and 150~GHz respectively \citep{sayre12, henning12}. 
The angular resolution at these frequencies is approximately $1\farcm7$ at 95\,GHz and $1\farcm2$ at 150\,GHz. 
From April 2013 through September 2016, the \sptpol{} receiver was used to survey a 500\,\sqdeg{} field. 
The field spans 15 degrees in declination (DEC) from $-65$ to $-50$~degrees and four hours in right ascension (RA) from $22^{\rm h}$ to $2^{\rm h}$. 
The final map noise levels in temperature are approximately 5.6\,\ukarcmin{} at 150\,GHz and 11.8\,\ukarcmin{} at 95\,GHz, in the multipole range $3000<\ell<5000$.

The time-ordered data (TOD) are bandpass filtered and coadded according to inverse noise variance weights into maps of the Stokes $I$, $Q$, and $U$ parameters. 
We use the flat-sky approximation, with a map pixel size of $0\farcm25$ in the Sanson-Flamsteed projection \citep{calabretta02, schaffer11}.
The map-making process is described in more detail in  \citet{crites15, keisler15, henning18}.

While bandpass filtering the TOD reduces the map noise levels, it also causes ringing around the location of unmasked, bright sources. 
This can bias the flux measurements of nearby sources.
This primarily happens in the scan direction, which is parallel to RA.  
One could mask all sources, but then the noise properties at the source locations might differ from the noise estimated at random locations, potentially affecting the noise bias correction (see Section~\ref{sec:noise_bias}).\footnote{In practice, we found 
nearly identical results ($<0.1\sigma$ difference in \mnpsq) when masking all sources as in the procedure described here, suggesting that any noise variation is negligible.}
Instead we create a set of maps with different sources masked in each map. 
The intent is to have each source unmasked for the measurement while masking any nearby source whose ringing might affect the main source.
We create two maps for measuring the polarisation of high flux ($S_{\rm 150\,GHz} \ge 40$\,mJy) sources, with all low flux sources masked. Each map contains one of the two disjoint sets of sources where the sets are defined by requiring the high-flux sources be separated by at least: $\Delta{\rm RA} \ge 100^\prime$ and $\Delta{\rm DEC} \ge 6^\prime$.
Similarly, we create two maps for low-flux sources ($S_{\rm 150\,GHz} < 40$\,mJy), requiring a small separation of  $\Delta{\rm RA} \ge 15^\prime$ and $\Delta{\rm DEC} \ge 6^\prime$. 
These separations are at least twice the distances where the ringing is negligible as confirmed by visual inspections of the maps at the source locations.

Relative gain errors between detectors and other instrumental non-idealities can leak total intensity (\I{}) into the polarisation maps. 
Specifically, this class of non-idealities leads to a monopole leakage where the temperature signal is mirrored in the Stokes \Q{}/\U{} maps. 
In the complete absence of monopole leakage, the mean \Q{} and \U{} signals of a large ensemble of point sources tend to zero due to their random polarisation angles.
Thus, we estimate this effect from the mean flux weighted \Q{}/\I{} and \U{}/\I{} signals of the point sources. 
We find monopole leakage factors of $\epsilon^Q=0.0182\pm0.0027$  and $\epsilon^U=-0.0095\pm0.0023$ at 95\,GHz and $\epsilon^Q=0.0015\pm0.0024$  and $\epsilon^U=0.0217\pm0.0028$ at 150\,GHz.
The error in these factors results in $5.8\times10^{-5}$ and $4.2\times 10^{-5}$ uncertainties in the mean squared polarisation fraction at 95 and 150\,GHz, respectively.
We subtract $\epsilon^Q I$  and $\epsilon^U I$ from the Stokes \Q{} and \U{} maps, respectively, and propagate the uncertainties to the measurements of the mean squared polarisation fraction.

Differential beam ellipticity between detector pairs will instead lead to a quadrupole leakage from temperature to Stokes \Q{}/\U{}. 
This differential beam ellipticity is measured to be on the order of 1~per cent using observations of Venus \citep{henning18}. 
The resulting leakage pattern is estimated from the Venus maps (assuming Venus is unpolarised).
The Venus-estimated leakage pattern is convolved by the \I{} map and subtracted from the \Q{}/\U{} maps. 
The final polarisation fraction results in this work are robust to errors on the measured quadrupole leakage; we have tested the extreme case of not subtracting the quadrupole leakage and do not find significant changes in the measured mean squared polarisation fractions. 

\section{Methods}
\label{sec:method}
We now describe the method to estimate the polarisation fraction of  AGN and DSFGs. 
We also test the performance of the estimator on simulations, and compare it to alternative schemes in the literature. 

The basic ingredients of the analysis are the Stokes \I{}/\Q{}/\U{} maps from \sptpol{} sampled at the locations of sources in the \sptsz{} source catalogue. 
These maps are apodized and zero-padded before applying a Fourier space matched filter for the point source that assumes white instrumental noise in the intensity maps. 
The same filter with the same level of white noise is used for all three (\I{}, \Q{}, and \U{}) maps. 
In effect this filter de-weights very large angular scales  (where the CMB and low-frequency noise is significant) and very small angular scales (where the instrumental beam has blurred out any signal). 
We take the \I{}/\Q{}/\U{} signal to be the value of the filtered \I{}/\Q{}/\U{} map at the source location. 
We do this independently for the 95 and 150\,GHz maps.

\subsection{Polarisation fraction}
\label{sec:pol_fraction}
The polarisation fraction of a source is defined as the ratio of the magnitude of the linear polarisation to the intensity signal. 
In terms of the Stokes \Q{} and \U{} parameters, the linear polarisation $P$, can be written as:
\BEA
\label{EQ:PQU}
P^2=Q^2+U^2.
\EEA
The square of the polarisation fraction $p^2$ is  defined by:
\BEA
\centering
\label{EQ:pol_frac}
p^2 \equiv \frac{P^2}{I^2} = \frac{Q^2 + U^2}{I^2}.
\EEA
A challenge in accurately estimating the polarisation fraction is that magnitudes are positive-definite quantities. 
Noisy estimates of the Stokes \I{}, \Q{}, and \U{} parameters introduce a systematic bias in the inferred polarisation fraction.
We discuss how this noise bias is handled in Section \ref{sec:noise_bias}.

Fig.~\ref{fig:stack} shows the stacked thumbnails of $I$ and $P$ SPTpol maps at the locations of the $S_{\rm 150\,GHz} \ge40$\,mJy sources. 
These images have not been corrected for the noise bias on $P$.
The dark ring around the stacked point sources is due to the matched filtering. 
The signal from the AGN is seen at very high signal-to-noise at 95 and 150\,GHz in both intensity and polarisation. 

\subsection{Noise bias correction}
\label{sec:noise_bias}

Magnitudes such as $P$ are naturally biased high by any noise in the estimate of \Q{} or \U{}. 
This bias becomes more significant at lower signal-to-noise. 
Handling this bias is thus critical to accurately measuring the polarisation fraction of faint sources. 
In the literature, various methods have been developed to de-bias the polarisation signal \citep[e.g.][and references therein]{wardle74}. 
However, \citet{simmons85} have compared various de-biasing methods and concluded that all of them leave some residual bias at low signal-to-noise. In a recent study, \cite{vidal16} have shown that residual bias is smaller at low signal-to-noise if an independent and high signal-to-noise measurement of polarisation angle is available. 

Fortunately, the goal of constraining the contribution of AGN and DSFGs to CMB polarisation power spectra needs accurate estimates of the polarised power, i.e.~\mnpsq{} not \mnp{}. 
 It is straightforward to estimate the contribution of the noise power in this case. 
We have a noisy estimate  $X^\prime$ of each Stokes parameter $X \in [I,Q,U]$, 
\BE
X^\prime = X +n_X,
\EE
 where $n_X$ is a Gaussian realisation of the noise power spectrum $N_X$. 
 Given the form of Equation~\ref{EQ:pol_frac}, we need an unbiased estimate of $X^2$ (again drawn from $X\in [I,Q,U]$), which can be constructed as:
 \BE
 X^2 = X^{\prime2} - N_X. 
 \EE
 We estimate the noise power $N_X$ for each Stokes map by taking the mean over random positions in the map, 
 \BE
 N_X = \left< X^2 \right>_{\rm random}.
 \EE
 The noise in the \sptpol{} maps varies with declination, so we estimate $N_X$ in ten different declination strips at 10,000 random locations. 
 Each source will fall into one of these ten strips. 
 To zeroth order, any remaining noise variation within the declination range of each strip will be cancelled by the real sources being roughly uniformly distributed within each strip. 

\subsection{Error bar estimation}
\label{sec:error_bars}
We use the bootstrapping technique with replacement to estimate the uncertainties on measured \mnpsq{} values. 
Note that we neglect uncertainty in the measurement of the noise bias as the number of sources in a set is always much smaller than the number of random positions being used to estimate the noise bias. 
For each set of $n_s$ sources, we have an associated set of $n_s$ \I{}, \Q{} and \U{} map cutouts from the matched filtered maps.  
We draw 10,000 random samples of $n_s$ map cutouts from this set with replacement.
For each of these 10,000 samples, we take the mean of the $n_s$ maps and determine the resulting \psq{} values. 
The standard deviation of these 10,000 mean \psq{} values is taken to be the uncertainty on the \mnpsq{} measurement for that set of sources.

\begin{figure*}
\centering
\includegraphics[scale=0.7]{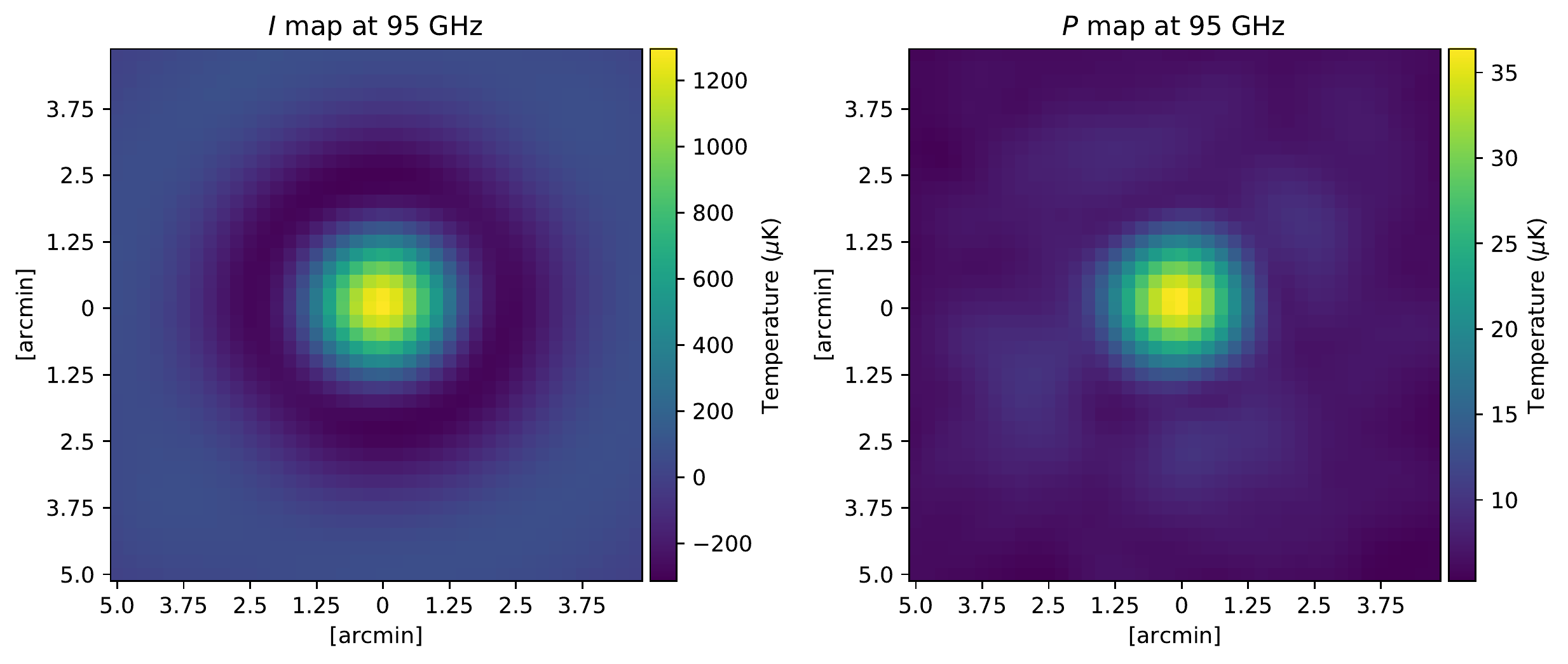}
\includegraphics[scale=0.7]{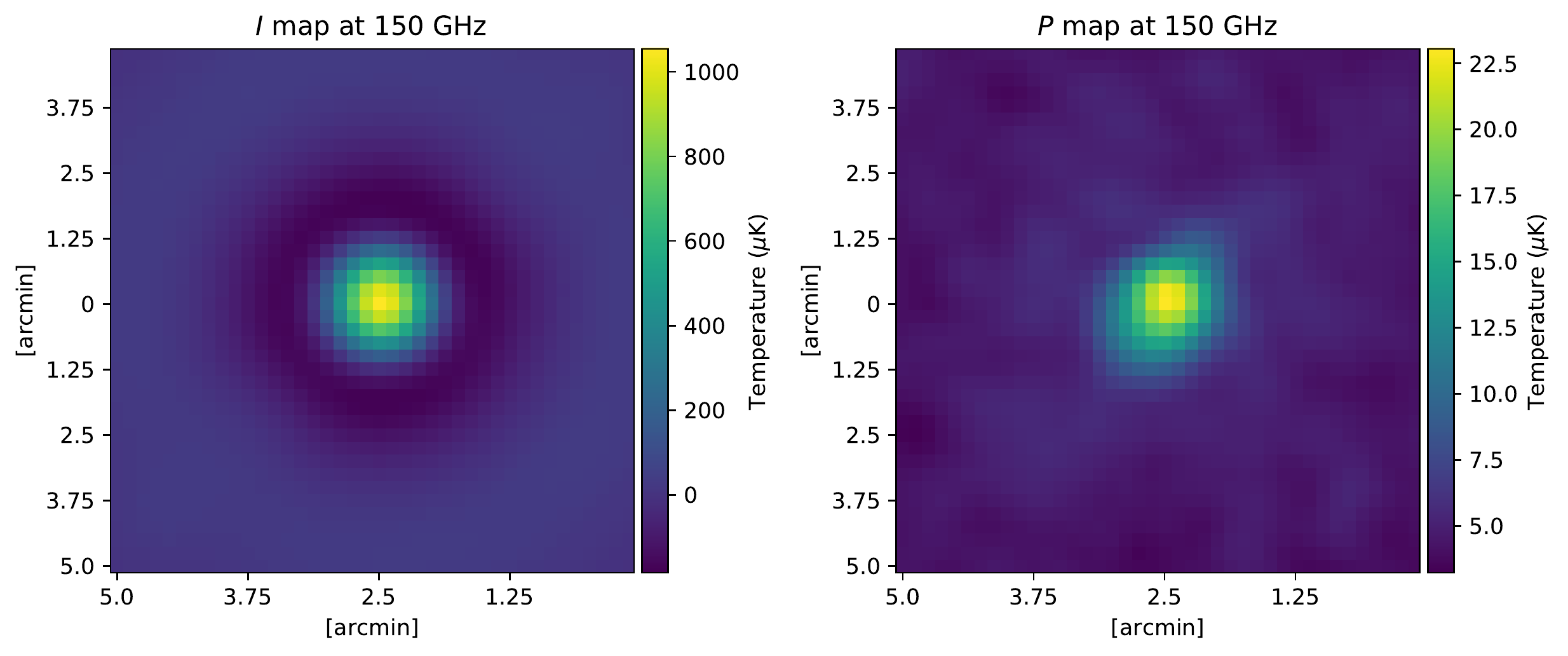}
\vskip-0.1in
\caption{Stacked $I$ and $P$ cutout maps extracted from the SPTpol 95\,GHz (top panels) and 150\,GHz (bottom panels) data at the positions of 69 sources with $S>40$~mJy in the SPT-SZ 150\,GHz catalogue described in Section~\ref{sec:catalog}.}
\label{fig:stack}
\end{figure*}
\begin{figure*}
\centering
\includegraphics[width=8.5cm, height=8cm]{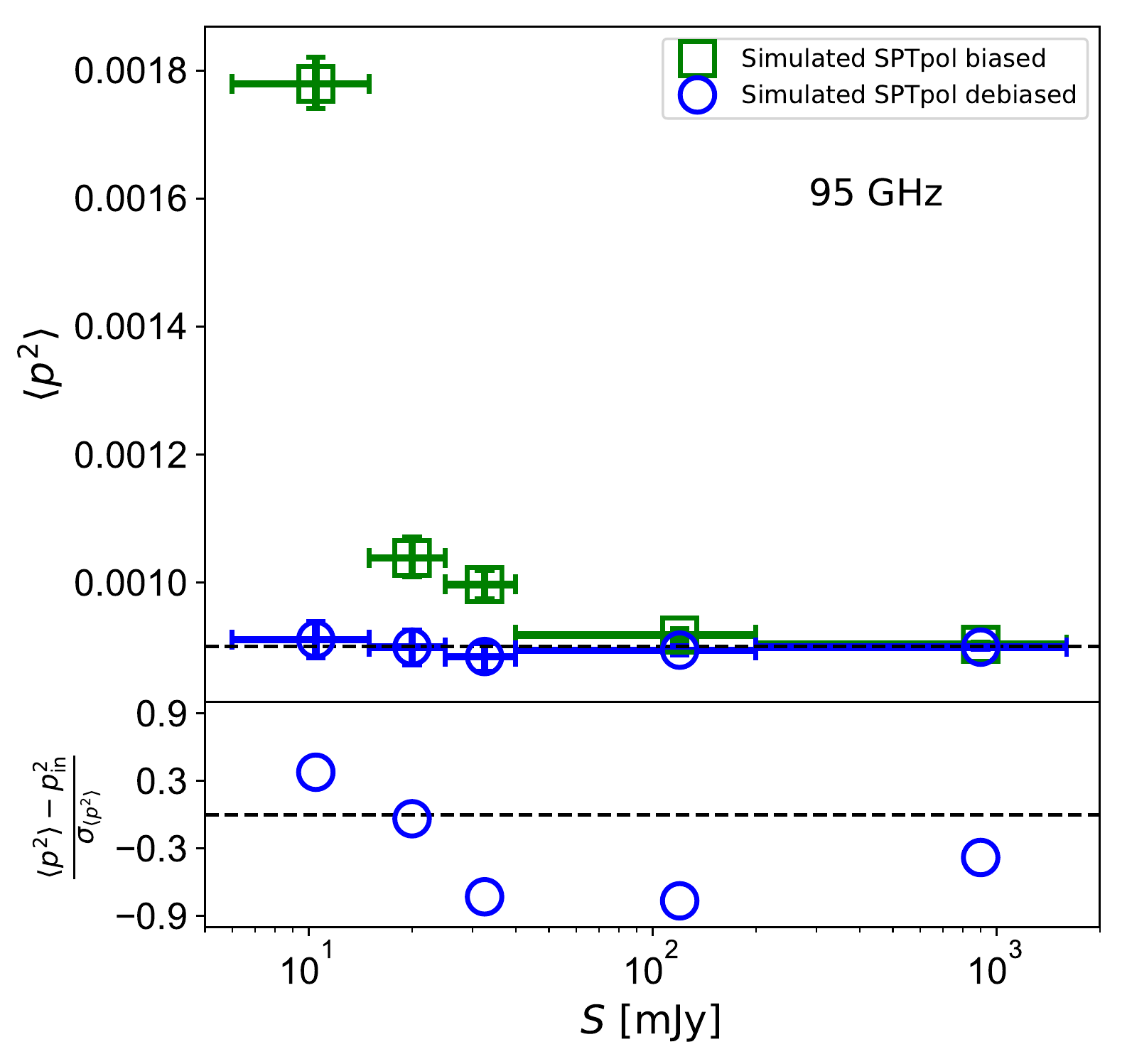}
\includegraphics[width=8.5cm, height=8cm]{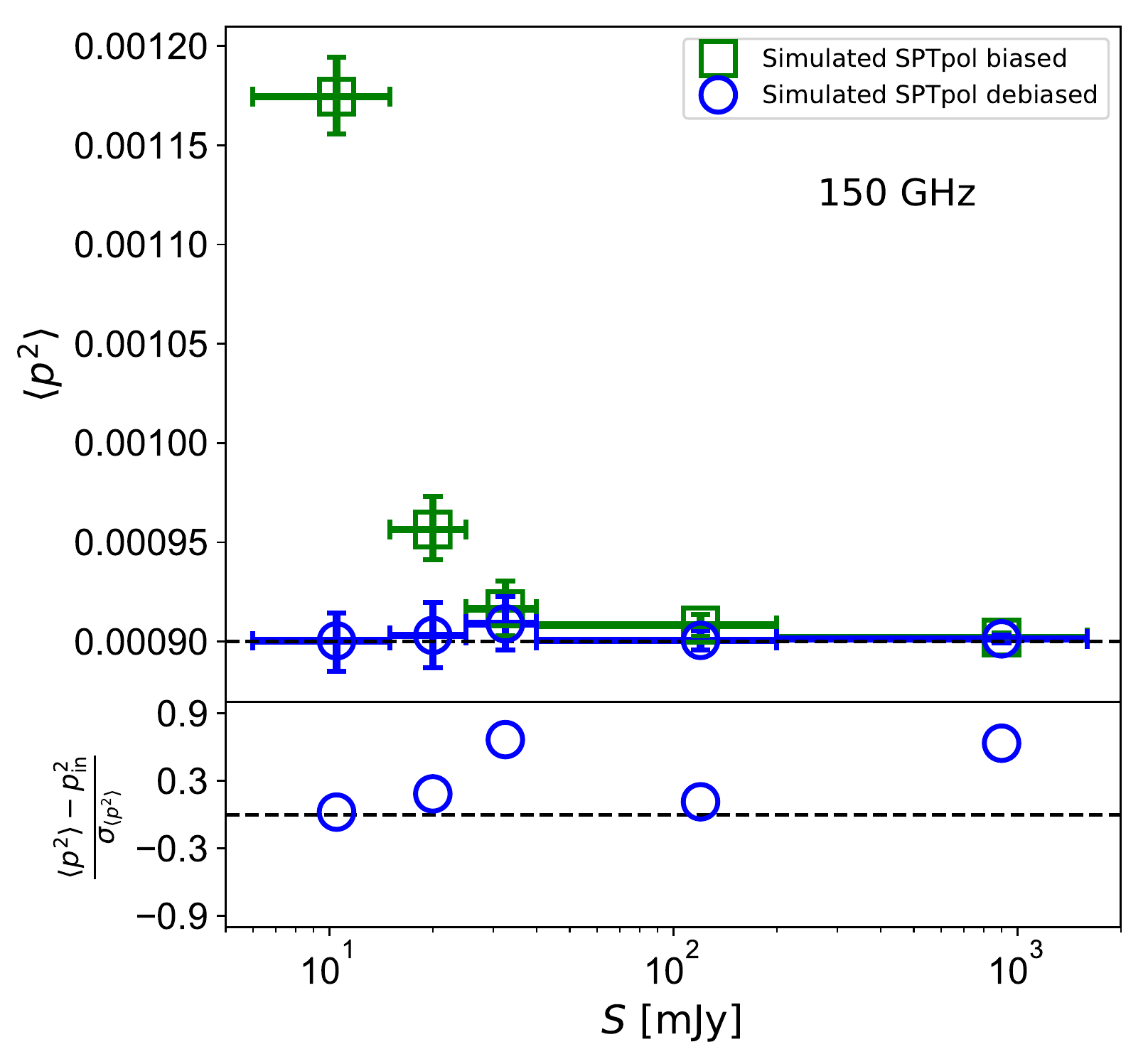}
\vskip-0.1in
\caption{  
Simulation results showing the recovered \mnpsq{} from the simulated maps at 95 (left panels) and 150\,GHz (right panels) where sources are injected with a constant input $p_{\rm in}^2=0.0009$ (solid black line) and random polarisation angles. 
We inject $\sim$ 7,000 sources to which flux is assigned using the measured $dN$/$dS$ distribution. 
The simulated maps are created from observed matched filtered SPTpol maps by masking all point sources which allows us to create simulated maps with same noise properties as in the observed maps. 
The recovered mean of the biased and de-biased $\langle p^2 \rangle$ values are shown as green and blue data points, respectively in different flux bins where the horizontal bars on data points represent bin size and the vertical error bars are computed using bootstrapping. 
The lower panels show the ratio of the difference between the recovered and input \psq{} value to the estimated uncertainty. 
The recovered value of \psq{} is consistent with the input in all flux bins within 1-$\sigma$ at both 95 and 150\,GHz. 
Given that the simulated sample is $10\times$ larger than the real source sample, this sets a 68 per cent CL upper limit on the magnitude of any bias to be $<0.35\sigma$ for the real data.}
\label{fig:simulation}
\end{figure*}
\subsection{Performance of the estimator}
\label{sec:simulations}
We test that we recover de-biased values for \mnpsq{} by injecting sets of simulated sources at random positions in the real matched filtered \sptpol{} \I, \Q, and \U{} maps. 
All known sources in the \sptsz{} catalogue with 150\,GHz flux $S_{\rm 150\,GHz} \ge6$\,mJy  are masked, and simulated sources are not injected into the masked areas. 
Seven thousand sources are injected, which is approximately ten times larger than the actual sample size.
The fluxes of the injected sources are drawn from the $dN$/$dS$ distribution of the SPT-SZ sources \citepalias{everett19}.
A random polarisation angle  is assigned to each source. 
All of the injected sources are taken to have the same polarisation fraction $p_{in} = 0.03 ~(p_{in}^2 = 0.0009)$. 
The injected sources are convolved by the \sptpol{} beam before being added to the real maps. 

As shown in Fig.~\ref{fig:simulation}, the \mnpsq{} estimator accurately recovers the input value, $p_{in}^2 = 0.0009$, within the statistical uncertainties of the simulated sample. 
The green points demonstrate the effect of the noise bias, which is much larger than the signal for objects with fluxes $\lesssim 40$\,mJy at 95 and 150\,GHz. 
However, even for the faintest objects at 95\,GHz where the noise bias can be two times larger  than the input polarisation signal, the de-biased estimate is consistent with the input value. 
As laid out in Section \ref{sec:error_bars}, uncertainties are estimated from the measured \psq{} distribution of the simulated sample. 
We note that the plotted uncertainties are for a sample that is approximately 10 times larger than the real sample, and thus the uncertainties are three times smaller than the real, noise-only uncertainty. 

To better visualize any residual bias in our estimator, the lower panel of Fig.~\ref{fig:simulation} shows the difference between the input and measured \psq{} values, in units of the 1-$\sigma$ uncertainties on the measurement in simulations. 
The agreement is excellent for all flux bins, showing no signs of systematic bias in the recovered polarisation fraction. 
Given that these simulated uncertainties are approximately $3\times$ smaller than the real uncertainties, this sets a 68 per cent CL upper limit that the magnitude of any bias is $<0.35\sigma$ for the real data.
We conclude that there is no evidence for bias in the measurement of the mean square polarisation fraction, \mnpsq{}.  
\begin{figure*}
\centering
\includegraphics[width=8.5cm, height=8cm]{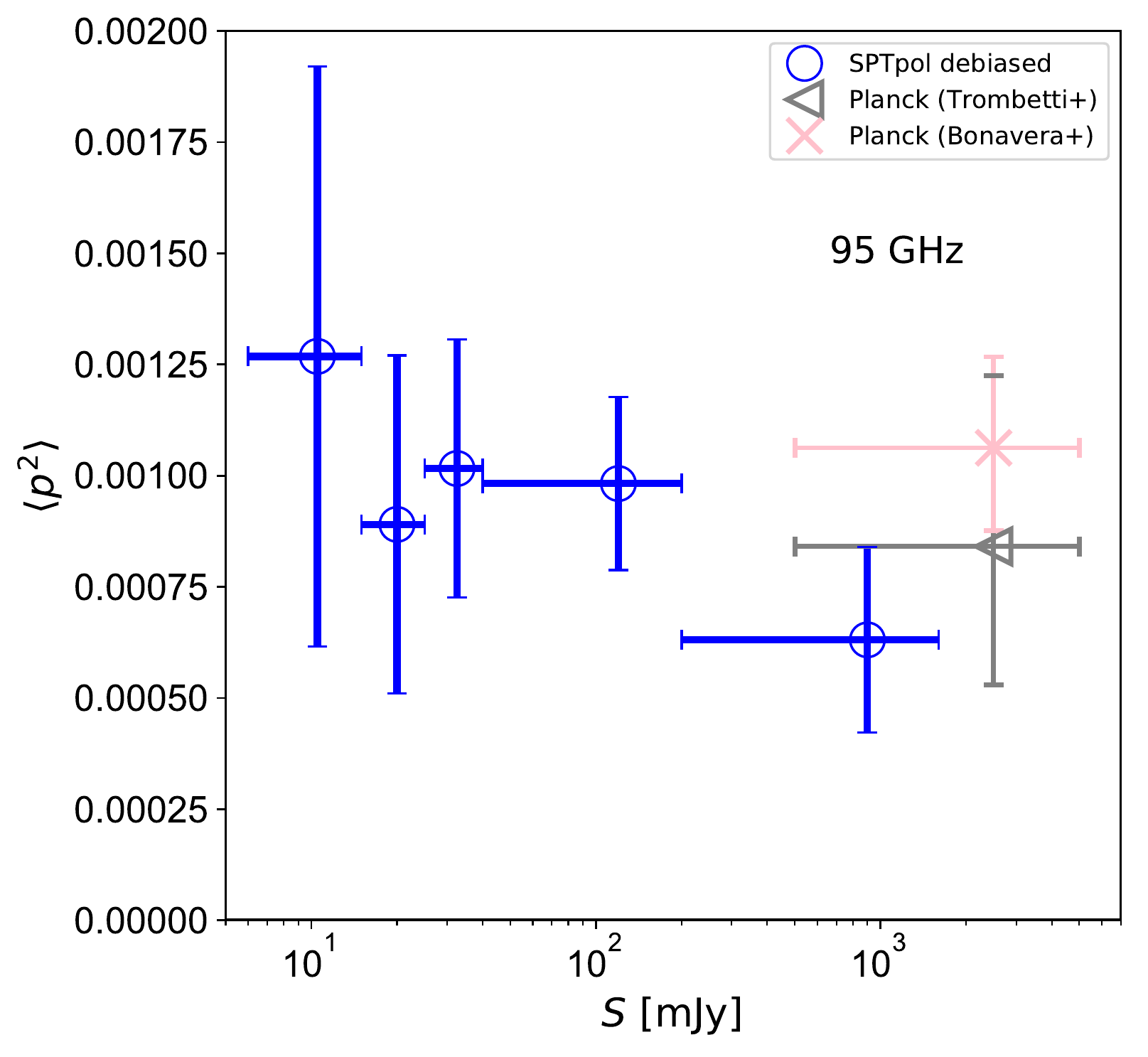}
\includegraphics[width=8.5cm, height=8cm]{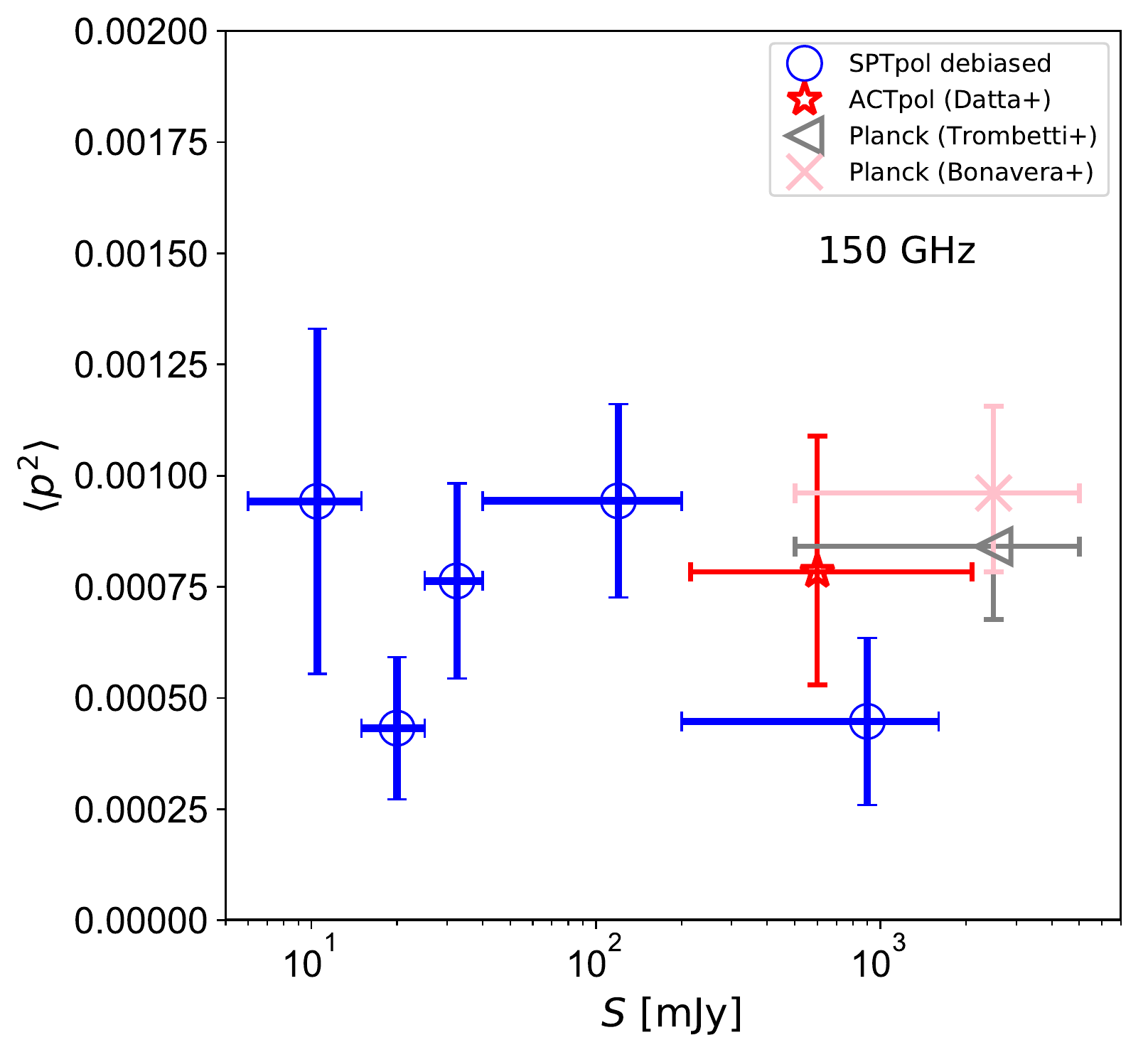}
\vskip-0.1in
\caption{Estimated $\mnpsq$ for extragalactic radio sources at 95 (left panel) and 150~GHz (right panel). 
The blue data points represent de-biased $\mnpsq$ values in five flux bins split according to the SPT-SZ 150 GHz flux in both panels. The horizontal bars on data points represent bin size and the vertical error bars are computed using bootstrapping (see Section~\ref{sec:error_bars}). 
For comparison, we also show the squared values of polarisation fraction $\langle p \rangle ^2$ measured by \citet{bonavera17}, \citet{trombetti18} both at 100 (left panel) and 143~GHz (right panel) and \citet{datta18} (at 148~GHz in the right panel) for high flux radio sources.}
\label{fig:SPTcat}
\end{figure*}
\begin{figure}
\centering
\includegraphics[width=0.4\textwidth]{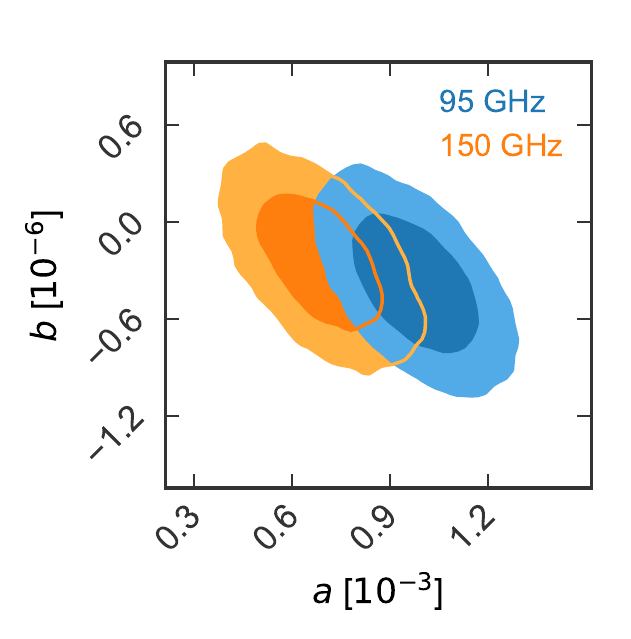}
\vskip-0.1in
\caption{
We find no evidence that the polarisation fraction depends on the flux of the source at either 95 or 150\,GHz. 
Here we show the 1 and 2-$\sigma$ contours from fitting the measured \mnpsq{} values to a linear function of flux, $\mnpsq = a+b\times S$, over the flux range $S \in[6,1500]$\,mJy. 
The slope is consistent with zero at both frequencies. 
The shape of the degeneracy between the offset and slope can be understood by the fact the polarisation fraction is well-measured for bright sources, but increasingly uncertain towards lower fluxes. 
The errors at 95~GHz are larger due to higher map noise levels (see Section \ref{sec:maps})}
\label{fig:contour}
\end{figure}
\begin{table}
\centering
\begin{center}
\begin{tabular}{lcccc}
\hline
\multicolumn{1}{c}{$S_{\rm range}$ $\rm[mJy]$} & \multicolumn{1}{c}{$N_{\rm source}$} & \multicolumn{1}{c}{$\langle p^2 \rangle$ [$10^{-4}$]} &  \multicolumn{1}{c}{$\sqrt{\langle p^2 \rangle}$ [$10^{-2}$]} \\ 
\hline
\multicolumn{4}{|c|}{95~GHz} \\
\hline
$6-15$          & $486$	   & $12.6\pm6.4$       & $3.56\pm0.82$ \\[3pt]
$15-25$        & $79$	       & $8.9\pm3.8$        & $2.98\pm0.58$ \\[3pt]
$25-40$       & $49$        & $10.2\pm2.9$        & $3.18\pm0.43$ \\[3pt]
$40-200$     & $53$        & $9.8\pm1.9$       & $3.13\pm0.31$ \\[3pt]
$200-1500$ & $15$         & $6.3\pm2.1$       & $2.51\pm0.40$ \\[3pt]
\\

\multicolumn{4}{|c|}{Unweighted Mean} \\
$6-1500$     & $682$       & $7.2\pm1.9$       & $2.68\pm0.36$ \\[3pt]
\multicolumn{4}{|c|}{Weighted Mean} \\
$6-1500$     & $682$        & $8.9\pm1.1$        &  $2.98\pm0.19$ \\[3pt]
\hline
\hline
\multicolumn{4}{|c|}{150~GHz} \\
\hline
$6-15$          & $490$	   & $9.4\pm3.9$      & $3.07\pm0.58$ \\[3pt]
$15-25$        & $79$	       & $4.3\pm1.5$        & $2.08\pm0.35$ \\[3pt]
$25-40$       & $49$        & $7.6\pm2.2$        & $2.76\pm0.37$ \\[3pt]
$40-200$     & $53$        & $9.4\pm2.1$       & $3.07\pm0.34$ \\[3pt]
$200-1500$ & $15$         & $4.5\pm1.8$       & $2.12\pm0.41$ \\[3pt]
\\
\multicolumn{4}{|c|}{Unweighted Mean} \\
$6-1500$     & $686$       & $5.3\pm1.7$       & $2.30\pm0.34$ \\[3pt]
\multicolumn{4}{|c|}{Weighted Mean} \\
$6-1500$     & $686$       & $6.9\pm1.1$        &  $2.63\pm0.22$ \\[3pt]
\hline
\end{tabular}
\end{center}
\caption{Mean-squared polarisation fractions ($\langle p^2 \rangle$) and inferred fractional polarisation ($\sqrt{\langle p^2 \rangle}$) measurements for $N_{\rm source}$ number of SPT-SZ selected sources stacked in flux bins $S_{\rm  range}$ using 95 and 150~GHz SPTpol maps. There are four sources missing at 95\,GHz due to smaller inverse noise variance weights at map edges. We show the unweighted (without any flux weighting) mean values for whole flux range as well. The error bars are evaluated using bootstrapping. We also show the flux weighted mean values, that is the best fit intercept, assuming zero slope for a straight line model fit to the mean-squared polarisation fraction in five flux bins (see Section~\ref{sec:p2_trends}).}
\label{Table:pol_fractions}
\end{table} 
\section{Results}
\label{sec:results}

Applying the methods of Section~\ref{sec:method} to the SPTpol maps at the SPT-SZ source locations yields a significant detection of the fractional source polarisation, as would be expected from Fig.~\ref{fig:stack}. 
When stacking the full source sample without any flux weighting, we find $\mnpsq = [7.2\pm1.9] \times 10^{-4}$ for 95\,GHz and $\mnpsq  = [5.3\pm1.7] \times 10^{-4}$ for 150\,GHz.
Weighting by flux yields a slight improvement in signal-to-noise. 
In this case, we find $\mnpsq = [8.9\pm1.1] \times 10^{-4}$ for 95\,GHz and $\mnpsq  = [6.9\pm1.1] \times 10^{-4}$ for 150\,GHz over the whole sample. 
The results are summarised in Table\,\ref{Table:pol_fractions}. 
Given the high significance detection, we now turn to considering possible trends with source properties.

\subsection{Trends with flux and observing frequency}
\label{sec:p2_trends}

An important question about the polarised foreground emission in CMB surveys is whether the polarisation fraction varies with flux or observing frequency. 
To answer these questions, we split the sample into five flux bins according to the SPT-SZ 150\,GHz flux, and measure the mean squared polarisation fraction in each bin. 
The results are listed in Table~\ref{Table:pol_fractions} and shown in Fig.~\ref{fig:SPTcat}. 
The horizontal bars in the figure represent the range of fluxes in each bin. 

Unsurprisingly, the uncertainties on the polarisation fraction increase towards lower fluxes even though the sample size increases in this direction. 
This trend is due to the uncertainty on the noise bias; effectively at lower fluxes, the estimator approaches the limit of subtracting two large numbers to find a small difference. 

To test whether the polarisation fraction depends on the source flux, we fit a straight line ($\mnpsq=a~+~b \times S$) to the measured \mnpsq{} across the five flux bins at both 95 and 150\,GHz. 
We use Markov Chain Monte Carlo (MCMC) and Gaussian likelihood for this.
The polarisation fraction measurements are assumed to be uncorrelated between different flux bins. 
Therefore, we model the covariance matrix as a diagonal matrix with entries given by the variance estimated through bootstrapping as discussed in Section~\ref{sec:error_bars}.
At 95 GHz, we find the best fit value for the offset is $a=[9.7\pm1.4] \times 10^{-4}$, with a slope of  $b=[-0.38\pm0.36] \times 10^{-6}$. 
The results at 150 GHz are similar: the preferred offset is $a=[6.9\pm1.3] \times 10^{-4}$ and the slope is $b=[-0.24\pm0.27] \times 10^{-6}$. 
We show both sets of parameter constraints in Fig~\ref{fig:contour}. 
Both slopes are within $1$-$\sigma$ of zero; the measured $\mnpsq$ does not show a statistically significant dependence on flux. 

We can also ask if the measured $\mnpsq$ varies with observing frequency from 95 to 150\,GHz. 
The measured offsets from the linear fits are consistent with no dependence on the observing frequency, with the best-fit values differing by only 1.5-$\sigma$.  
We can reduce these uncertainties by fitting a constant (i.e. fixing $b=0$) to the five flux bins. 
We find best-fit offsets of $a=[8.9\pm1.1] \times 10^{-4}$ at 95\,GHz and $a=[6.9\pm1.1] \times 10^{-4}$ at 150\,GHz. 
As shown in Table~\ref{Table:pol_fractions}, these values are consistent with (and have slightly smaller error bars than) the results for the unweighted (without any flux weighting) stack of the full sample.
The reduction in uncertainty can be understood by the weighting: in all stacks in this work, every source in the stack has equal weight. 
Splitting the data by flux bins weights the higher signal-to-noise high-flux sources more heavily. 
Given the best-fit offsets differ by only 1.3\,$\sigma$, we conclude that the data are consistent with the hypothesis that the polarisation fraction is the same at both frequencies. 

\subsection{Radio and dusty sources}
\label{sec:radio_dusty}

As mentioned in Section~\ref{sec:intro}, extragalactic sources that are bright at CMB frequencies are classified into two categories: radio sources (AGN) and dusty sources (DSFGs).
Although both populations affect the measurements of CMB E \& B modes of polarisation, it is interesting to compare the polarisation properties of these classes of sources separately. 
As described in Section~\ref{sec:catalog}, approximately 92 per cent of point sources with $S_{150}\geq 6$\,mJy observed in SPT-SZ survey are AGN, thus most of the signal in the stack of the whole sample is coming from them. 
Stacking just radio sources we find unweighted $\mnpsq = [7.3\pm2.0] \times 10^{-4}$ and $[5.2\pm1.8] \times 10^{-4}$ at 95 and 150\,GHz, respectively.
The uncertainties are the same as for the full sample as most of the DSFGs in the sample are low flux sources.
Similar to Section~\ref{sec:p2_trends}, fitting a constant (with slope fixed to zero) to this sub-sample of radio sources in five flux bins gives flux weighted $\mnpsq=[8.9\pm1.1] \times 10^{-4}$ and $[6.9\pm1.1] \times 10^{-4}$ at 95\,GHz and 150\,GHz, respectively.
These unweighted and weighted numbers are consistent with the values obtained from whole sample as listed in Table~\ref{Table:pol_fractions}.

The \citetalias{everett19} sample includes 55 sources classified as dusty sources in SPTpol observing region with $S_{150}\geq 6$\,mJy,  based on the spectral index from 150 to 220\,GHz. 
All of these dusty sources have 150\,GHz flux below $S=35$\,mJy, and 93 per cent of them are in the lowest flux bin of this work ($S$\,$\leq 15$\,mJy). 
For dusty sources we find unweighted $\mnpsq$ to be consistent with zero with large error bars i.e. $\mnpsq = [51\pm59] \times 10^{-4}$ and $[8.1\pm9.2] \times 10^{-4}$ at 95 and 150\,GHz, respectively. 
The resulting 95 per cent confidence level upper limits are $\mnpsq_{\rm 95}<16.9 \times 10^{-3}$ and $\mnpsq_{\rm 150}<2.6 \times 10^{-3}$.

\subsection{Comparison to previous results}
\label{sec:p2_previous}

The results shown in Table~\ref{Table:pol_fractions} are consistent with previous studies of AGN and DSFGs. 
Briefly, in previous studies, polarisation fraction is found to be independent of flux \citep{battye11, trombetti18, datta18} and frequency \citep{battye11,galluzzi17,bonavera17,galluzzi18,trombetti18} at GHz frequencies, and overall mean $\langle p \rangle$ is estimated at a level around $1-5$ per cent, which are all consistent with our findings in this work. 
For comparison, we show the squared values of polarisation fraction $\langle p \rangle ^2$  from \cite{bonavera17} and \citet{trombetti18} (at 100 and 143~GHz {\it Planck} frequencies) in both panels of Fig~\ref{fig:SPTcat}. 
In the right panel, we also show $\langle p \rangle ^2$ measured by \cite{datta18} at 148~GHz ACTpol observing frequency. 
Even though, these previous studies at high frequencies were performed for the brightest sources observed in separate CMB polarisation experiments and by using different estimators for noise bias correction, we find $\langle p \rangle ^2$ estimated in these studies in good agreement with our \mnpsq ~measurements at both frequencies.

On the dusty side, \citet{bonavera17a} used \planck{} data to study $\sim 4700$ dusty sources selected at 857~GHz with $S\geq 791$\,mJy (a flux threshold that is approximately comparable to this work when extrapolated to 150\,GHz). 
For these dusty sources, they found $3.52\pm 2.48$, $3.10\pm 0.75$ and $3.65\pm 0.66$ per cent polarisation fraction at 143, 217 and 353\,GHz, respectively. 
Using a different method and sources selected at 353\,GHz with $S\geq 784$\,mJy, \citet{trombetti18}  set 90 per cent CL upper limits on the polarisation fraction of $\mnp < 0.039$ at 217 GHz and $\mnp < 0.022$ at 353 GHz. 
Comparing these with our measurements at 95 and 150\,GHz from the last section, we see a trend of decreasing upper limits on the polarisation fraction with increasing frequency, possibly due to larger signal-to-noise of dusty sources at higher frequencies.
The upper limits on dusty sources at 150\,GHz from the last section are compatible with the \citet{bonavera17a}  measurements. 

\begin{figure*}
\centering
\includegraphics[width=0.49\textwidth]{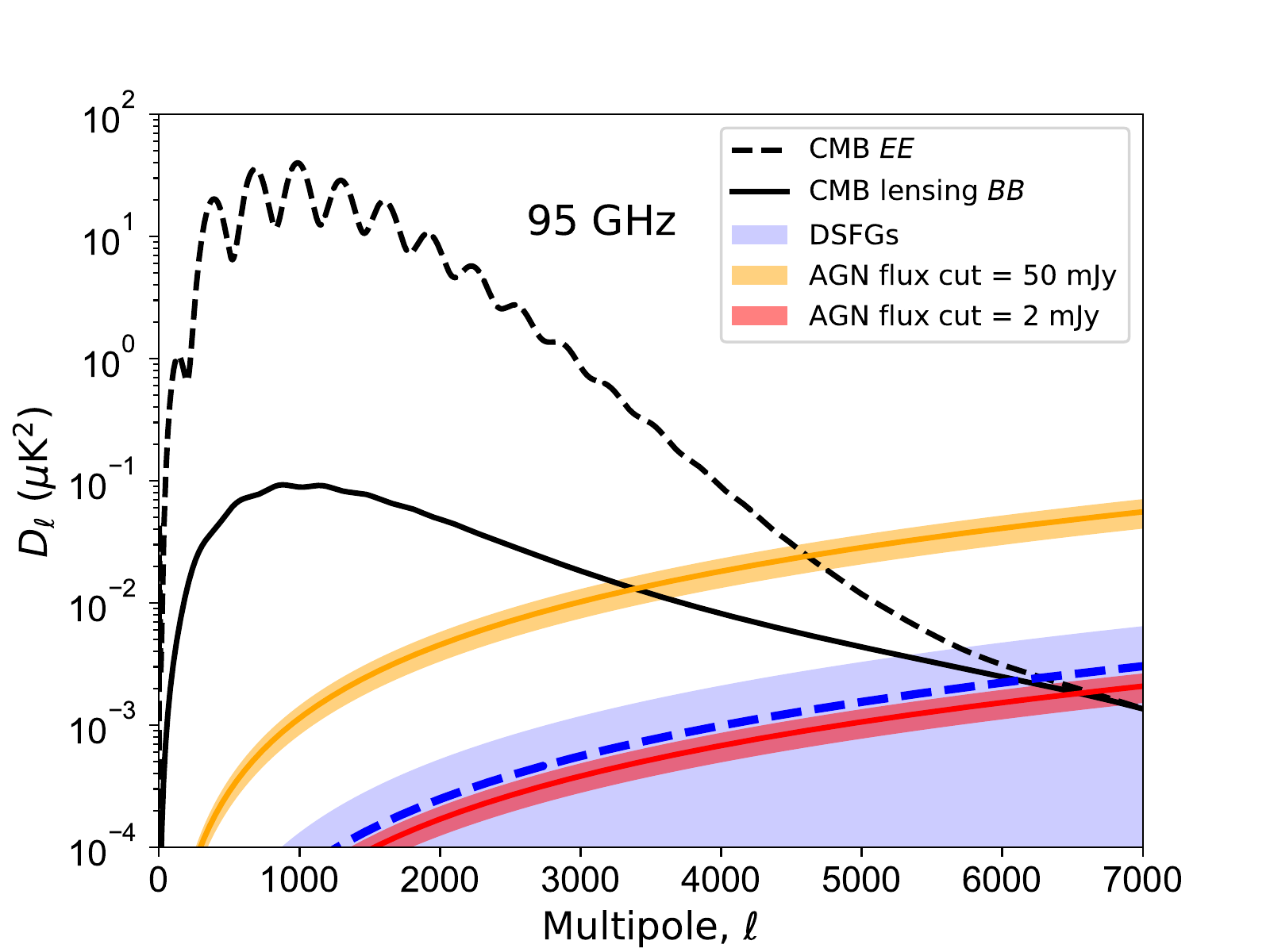}
\includegraphics[width=0.49\textwidth]{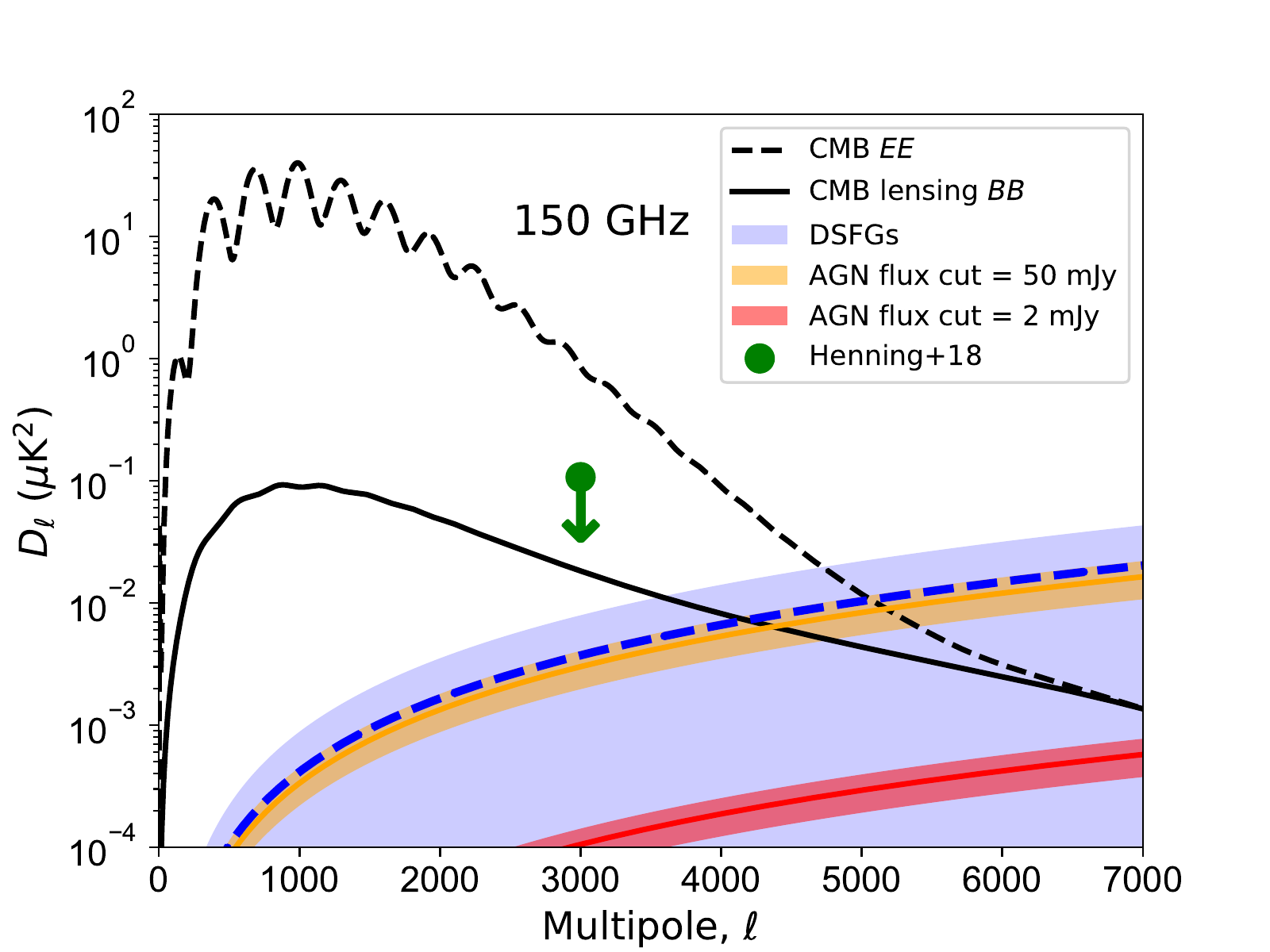}
\vskip-0.1in
\caption{
CMB polarisation surveys will be minimally impacted by polarised source power after masking radio and dusty galaxies above reasonable flux thresholds. 
The predicted AGN power at 95\,GHz (left panel) and 150\,GHz (right panel) for the \mnpsq{} measured in this work is shown by the solid red (yellow) line for a 50 (2) mJy masking flux threshold in temperature. 
The filled area represents the 1-$\sigma$ uncertainties on \mnpsq{} (no uncertainty in the source distribution has been included). 
The dashed blue line and the filled blue area show the mean and the 1-$\sigma$ upper limit on the predicted DSFG polarised power, respectively.
We have assumed that the fractional polarisation of the DSFGs remains constant from the value measured at 150\,GHz down to 95\,GHz.
The total polarised foreground power will be less than the CMB EE power spectrum out to $\ell\lesssim5700$ ($\ell\lesssim4700$) and CMB lensing BB power spectrum out to $\ell\lesssim5300$ ($\ell\lesssim3600$) at 95 (150)\,GHz.
The green arrow in the right panel shows the 95~per cent CL upper limit from the recent SPTpol EE power spectrum measurement \citep{henning18}. 
The polarisation fraction measurements in this work support the expectation that extragalactic foregrounds will be fractionally smaller for CMB polarisation than temperature measurements, thus allowing more modes to be used in polarisation analyses.
}
\label{fig:power}
\end{figure*}

\section{Implications for CMB measurements}
\label{sec:power_spectrum}

Polarised emission from AGN and DSFGs can contaminate measurements of CMB E \& B modes on small angular scales (high-$\ell$). 
The polarised power from these objects, especially in the case of AGN, can be substantially reduced by masking the brightest sources. 
However, there are practical limits on how many sources can be masked, both because of the detection threshold in an experiment and because 
the fraction of the map that is masked naturally rises as the mask flux threshold is lowered. 
Thus some residual polarised power from AGN and DSFGs  will always remain in measurements of the CMB E and B mode power spectra.
The measurement of the mean squared polarisation fraction in this work provides a direct means to predict this residual polarised power as a function of the masking threshold.

The power contribution to E and B modes from extragalactic sources will be equal on average given that the polarisation angles of point sources are distributed randomly. 
The point source power $C_{\ell, \rm PS}$ contribution to E and B modes is then given as
\BEA
\label{EQ:power1}
C_{\ell, \rm PS}^{EE} = C_{\ell, \rm PS}^{BB} = \frac{1}{2} \langle p^2 \rangle C_{\ell, \rm PS}^{TT},
\EEA
where $\langle p^2 \rangle $ is the unweighted mean-squared polarisation fraction for either the AGN or DSFG samples (see Section~\ref{sec:radio_dusty}) and $C_{\ell, \rm PS}^{TT}$ is the temperature power spectrum which will equal a constant for a spatially invariant Poisson distribution. 
The polarised clustered power is expected to be suppressed due to the random polarisation angles to negligible levels. 
For DSFGs, we take the measured central values of the Poisson power, $D_{\rm 3000}^{\rm 95\,GHz} = 1.37\,\uksq$ and $D_{\rm 3000}^{\rm 150\,GHz} = 9.16\,\uksq$, for the baseline model in \citet{george15}. 
These power levels were measured for a flux masking threshold of 6.4\,mJy at 150\,GHz; we neglect any variation with masking threshold since (i) almost all DSFGs have fluxes below 2\,mJy and (ii) the uncertainty on the polarisation fraction is large.

For AGN, we calculate the expected $C_{\ell}^{TT}$ as a function of the masking threshold according to the source flux distribution  $dN/dS$ of the C2Ex model  \citep{tucci11} at 150\,GHz. 
The C2Ex model builds on earlier models by e.g.~\cite{dezotti05} and is constructed by extrapolating the differential number counts of extragalactic sources observed at $\sim 5$\,GHz \cite[see][]{dezotti10} to higher frequencies.
\citet{tucci11} also compare the modeled number counts to the observed number counts in the SPT \citep{vieira10}, the ACT \citep{marriage11a}, and the \textit{Planck} \citep{planck11-7} surveys. 
 \cite{mocanu13} compared the C2Ex model to observed point source counts in 720\,\sqdeg{} SPT-SZ survey and found it consistent with sources above 80\,mJy and below 20\,mJy. 
The C2Ex model predicts $D_{\ell =3000, \rm PS}^{TT} =  0.4 \,(11.4)\,\uksq$ for a masking threshold of 2 (50)\,mJy at 150\,GHz. 
We scale these powers to 95\,GHz using a spectral index of -0.9, which is the value preferred in \citet{george15}. 
This predicts radio power of $D_{\ell =3000, \rm PS}^{TT} =  1.1 \,(28.6)\,\uksq$ for the same masking thresholds: 2 (50)\,mJy.

Fig.~\ref{fig:power} shows the predicted power spectrum ($D_{\ell} = \ell (\ell+1) C_{\ell} / 2\pi$) at 95\,GHz (left panel) and 150\,GHz (right panel). 
The plots show the cosmological CMB EE and BB power spectra as well as the polarised power predicted for AGN (at two possible flux cut limits) and DSFGs, given the polarisation fractions measured in this work. 
These figures illustrate that the polarised power from AGN and DSFGs will significantly contaminate measurements of the cosmological EE and BB power spectra only on the smallest angular scales. 
The main uncertainty in both panels is the polarisation fraction of the DSFGs. 
We use the DSFG \mnpsq{} measured at 150\,GHz for both 95 and 150\,GHz. 
Given that both frequencies are in the extreme Rayleigh tail of the dust black body spectrum, we do not expect a significant shift in the sources probed or in the polarisation of each source between 95 and 150\,GHz. 
Supporting this position, \citet{bonavera17a} found  the DSFG $\langle p \rangle$  to not vary with frequency from 143 to 353\,GHz. 
For a reasonable mask flux threshold of 2\,mJy which is achievable by existing experiments like SPT-3G, the total polarised foreground power is less than the EE power spectrum out to $\ell\lesssim5700$ ($\ell\lesssim4700$) and the BB lensing power spectrum out to $\ell\lesssim5300$ ($\ell\lesssim3600$) at 95 (150)\,GHz.

The equivalent angular multipole for temperature is $\ell \sim 4000$  and $\ell \sim 3100$ at 95 and 150\,GHz, respectively -- more modes will be available to cosmological studies in polarisation than temperature.

Note that the results are not particularly sensitive to the flux cut since the polarised foreground power envelope is being driven by DSFGs. 
The DSFG intensity power varies slowly with mask threshold since most DSFGs are fainter than 2\,mJy. 
It is also worth noting that better measurements of the DSFG \mnpsq{} are likely to make the AGN power, and thus masking threshold, somewhat more important. 
We have reasons to believe that the polarisation of DSFGs is lower than AGN, but 1-$\sigma$ limits in this work are higher (at 150\,GHz) for DSFGs due to the limited number of DSFGs in the sample. 

The inferred power from the measured \mnpsq{} values in this work are compatible with current observational constraints. 
Using the measured EE bandpowers from the SPTpol survey at 150\,GHz, \cite{henning18} places a 95 per cent confidence level upper limit of $D_{\ell=3000}^{\rm PS}< 0.107$ $\mu \rm K^2$ on polarised point source contribution to the EE power spectrum. 
This result is for a flux mask threshold of 50\,mJy. 
The predicted power in our analysis  is $D_{\ell=3000}^{\rm PS} < 0.0092$ $\mu \rm K^2$ at 150\,GHz, well below the observed upper limit. 
The other polarisation fraction measurements discussed in Section~\ref{sec:p2_previous} such as \citet{datta18} also argue for similar, low levels of polarised foreground power.

\section{Conclusions}
\label{sec:conclusions}
We present a new method to measure the mean squared polarisation fraction of sources in CMB surveys. 
We apply the method to 95 and 150\,GHz maps from the \sptpol{} 500\,\sqdeg{} survey at the locations of sources selected to have $S_{150}\geq 6$~mJy  in the \sptsz{} survey, and find \mnpsq{} for five flux bins. 
The flux-weighted mean squared polarisation fraction i.e. the best fit value across the five bins is $\mnpsq_{\rm 95} = [8.9\pm1.1] \times 10^{-4}$  and $\mnpsq_{\rm 150}  = [6.9\pm1.1] \times 10^{-4}$.
We find no evidence in the current data that the polarisation fraction varies with observing frequency or source flux density. 

We split the source sample into DSFGs and AGN based on the observed spectral index from 150 to 220\,GHz \citep{everett19}. 
At 150\,GHz, we find $\mnpsq_{\rm AGN} =  [5.3\pm1.7] \times 10^{-4}$ and $\mnpsq_{\rm DSFG} =  [8.1\pm9.2] \times 10^{-4}$ without any flux weighting.
The larger uncertainties for the DSFG sample are due to the limited number of DSFGs above 6\,mJy at 150\,GHz. 
We use these measured mean squared polarisation fractions to predict the extragalactic foreground contribution to the CMB polarisation power spectra.

Given the 1-sigma upper limit on the \mnpsq{} measured at 150\,GHz in this work, the extragalatic foreground power will be subdominant to the CMB E mode power spectrum for $\ell\lesssim5700$ ($\ell\lesssim4700$) and to the CMB B-mode power spectrum for $\ell\lesssim5300$ ($\ell\lesssim3600$)  at 95 (150)\,GHz.

These are lower limits on angular multipoles and most likely the CMB polarisation power spectra will dominate out to even higher multipoles. 
In comparison, these extragalactic foregrounds surpass the CMB temperature power spectrum around $\ell \sim 4000$  and $\ell \sim 3100$ at 95 and 150\,GHz, respectively for the same flux cuts.
With the exquisitely low noise levels expected for current and upcoming experiments like SPT-3G \citep{bender18} and CMB-S4 \citep{abazajian16}, we will thus be able to recover more cosmological information from CMB polarisation anisotropies than temperature anisotropies by virtue of going to much smaller angular scales.

\section*{Acknowledgements}
SPT is supported by the National Science Foundation through grant PLR-1248097.  Partial support is also provided by the NSF Physics Frontier Center grant PHY-1125897 to the Kavli Institute of Cosmological Physics at the University of Chicago, the Kavli Foundation and the Gordon and Betty Moore Foundation grant GBMF 947. This research used resources of the National Energy Research Scientific Computing Center (NERSC), a DOE Office of Science User Facility supported by the Office of Science of the U.S. Department of Energy under Contract No. DE-AC02-05CH11231.  
The Melbourne group acknowledges support from the Australian Research Council's Discovery Projects scheme (DP150103208). 

\bibliographystyle{mnras}
\bibliography{PS_polarisation}

\section*{Author Affiliations}
\Melbourne School of Physics, University of Melbourne, Parkville, VIC 3010, Australia \\
\Cardiff Cardiff University, Cardiff CF10 3XQ, United Kingdom\\
\FNAL Fermi National Accelerator Laboratory, MS209, P.O. Box 500, Batavia, IL 60510\\
\KICPChicago Kavli Institute for Cosmological Physics, University of Chicago, 5640 South Ellis Avenue, Chicago, IL, USA 60637 \\
\illast Astronomy Department, University of Illinois at Urbana-Champaign, 1002 W. Green Street, Urbana, IL, USA 61801\\
\NIST NIST Quantum Devices Group, 325 Broadway Mailcode 817.03, Boulder, CO, USA 80305\\
\Berkeley Department of Physics, University of California, Berkeley, CA, USA 94720\\
\ArgonneHEP High Energy Physics Division, Argonne National Laboratory, 9700 S. Cass Avenue, Argonne, IL, USA 60439\\
\AAUChicago Department of Astronomy and Astrophysics, University of Chicago, 5640 South Ellis Avenue, Chicago, IL, USA 60637\\
\PhysicsUChicago Department of Physics, University of Chicago, 5640 South Ellis Avenue, Chicago, IL, USA 60637\\
\EFIChicago Enrico Fermi Institute, University of Chicago, 5640 South Ellis Avenue, Chicago, IL, USA 60637\\
\UKZN School of Mathematics, Statistics \& Computer Science, University of KwaZulu-Natal, Durban, South Africa\\
\UChicago University of Chicago, 5640 South Ellis Avenue, Chicago, IL, USA 60637\\
\TAPIRCaltech TAPIR, Walter Burke Institute for Theoretical Physics, California Institute of Technology, 1200 E California Blvd, Pasadena, CA, USA 91125\\
\Caltech California Institute of Technology, MS 249-17, 1216 E. California Blvd., Pasadena, CA, USA 91125\\
\LBNL Physics Division, Lawrence Berkeley National Laboratory, Berkeley, CA, USA 94720\\
\McGill Department of Physics, McGill University, 3600 Rue University, Montreal, Quebec H3A 2T8, Canada\\
\CIFAR Canadian Institute for Advanced Research, CIFAR Program in Cosmology and Gravity, Toronto, ON, M5G 1Z8, Canada\\
\ColoradoAPS Department of Astrophysical and Planetary Sciences, University of Colorado, Boulder, CO, USA 80309\\
\illphy Department of Physics, University of Illinois Urbana-Champaign, 1110 W. Green Street, Urbana, IL, USA 61801\\
\HarveyMudd Harvey Mudd College, 301 Platt Blvd., Claremont, CA 91711\\
\esogarching European Southern Observatory, Karl-Schwarzschild-Str. 2, 85748 Garching bei M\"{u}nchen, Germany\\
\ColoradoPhys Department of Physics, University of Colorado, Boulder, CO, USA 80309\\
\SLAC SLAC National Accelerator Laboratory, 2575 Sand Hill Road, Menlo Park, CA 94025\\
\Stanford Dept. of Physics, Stanford University, 382 Via Pueblo Mall, Stanford, CA 94305\\
\Davis Department of Physics, University of California, One Shields Avenue, Davis, CA, USA 95616\\
\Arizona Steward Observatory, University of Arizona, 933 North Cherry Avenue, Tucson, AZ 85721\\
\Michigan Department of Physics, University of Michigan, 450 Church Street, Ann  Arbor, MI, USA 48109\\
\Munich Faculty of Physics, Ludwig-Maximilians-Universit\"{a}t, Scheinerstr.\ 1, 81679 Munich, Germany \\
\ExcellenceCluster Excellence Cluster Origins, Boltzmannstr.\ 2, 85748 Garching, Germany \\
\MPE Max Planck Institute for Extraterrestrial Physics, Giessenbachstr.\ 85748 Garching, Germany \\
\Dunlap Dunlap Institute for Astronomy \& Astrophysics, University of Toronto, 50 St George St, Toronto, ON, M5S 3H4, Canada\\
\ArgonneMSD Materials Sciences Division, Argonne National Laboratory, 9700 S. Cass Avenue, Argonne, IL, USA 60439\\
\Minnesota School of Physics and Astronomy, University of Minnesota, 116 Church Street S.E. Minneapolis, MN, USA 55455\\
\CaseWestern Physics Department, Center for Education and Research in Cosmology and Astrophysics, Case Western Reserve University, Cleveland, OH, USA 44106\\
\ArtInstChicago Liberal Arts Department, School of the Art Institute of Chicago, 112 S Michigan Ave, Chicago, IL, USA 60603\\
\ThreeSpeedLogic Three-Speed Logic, Inc., Vancouver, B.C., V6A 2J8, Canada\\
\JPL Jet Propulsion Laboratory, California Institute of Technology, Pasadena, CA. USA 91109\\
\CfA Harvard-Smithsonian Center for Astrophysics, 60 Garden Street, Cambridge, MA, USA 02138\\
\KIPAC Kavli Institute for Particle Astrophysics and Cosmology, Stanford University, 452 Lomita Mall, Stanford, CA 94305\\
\GSFC NASA Goddard Space Flight Center, Greenbelt, MD 20771\\
\UToronto Department of Astronomy \& Astrophysics, University of Toronto, 50 St George St, Toronto, ON, M5S 3H4, Canada\\
\Maryland Department of Astronomy, University of Maryland College Park, MD, USA 20742\\
\UCLA Department of Physics and Astronomy, University of California, Los Angeles, CA, USA 90095\\

\end{document}